\documentclass[12pt]{book}

\usepackage[dvips]{graphicx,color}
\usepackage{makeidx,view}

\makeauthorindex
\makeindex

\begin{document}

\BookTitle{\itshape The Universe Viewed in Gamma-Rays}
\CopyRight{\copyright 2002 by Universal Academy Press, Inc.}
\pagenumbering{arabic}

\chapter{STACEE Observations of Mrk 421 During the 2001 Flare}

\author{%
        L.~M. Boone$^1$,
        D. Bramel$^3$,
        E. Chae$^2$,
        C.~E. Covault$^4$,
        P. Fortin$^5$,
        D.~M. Gingrich$^{6,7}$,
        D.~S. Hanna$^5$,
        J.~A. Hinton$^{2,11}$,
        R. Mukherjee$^3$,
        C. Mueller$^5$,
        R.~A. Ong$^8$,
        K. Ragan$^5$,
        R.~A. Scalzo$^2$,
        D.~R. Schuette$^{8,9}$,
        C.~G. Th\'eoret$^{5,10}$,
 	and D.~A. Williams$^1$\\
{\it
	$^1$Santa Cruz Institute for Particle Physics, Santa Cruz, CA 95064, USA;
	$^2$Enrico Fermi Institute, University of Chicago, 5640 Ellis Ave., Chicago, IL 60637, USA;
	$^3$Columbia University \& Barnard College, New York, NY 10027, USA;
	$^4$Department of Physics, Case Western Reserve University, 10900 Euclid Ave., Cleveland, OH 44106, USA;
	$^5$Department of Physics, McGill University, Montreal, Quebec H3A 2T8, Canada;
	$^6$Centre for Subatomic Research, University of Alberta, Edmonton, Alberta T6G 2N5, Canada;
	$^7$TRIUMF, Vancouver, British Columbia V6T 2A3, Canada;
	$^8$Department of Physics \& Astronomy, University of California, Los Angeles, CA 90095, USA;
	$^9$Present address: Department of Physics, Cornell University, Ithaca, NY 14853;
	$^{10}$Present address: Laboratoire de Physique Corpusculaire et Cosmologie, Coll\`ege de France,
		F-75231 Paris CEDEX 05, France;
	$^{11}$Present address: Max-Planck-Institut f\"ur Kernphysik Postfach 10 39 80, D-69029 Heidelberg, Germany
}
}

%
\AuthorContents{L.\ M.\ Boone et al.} 

\AuthorIndex{Boone}{L.\ M.}
\AuthorIndex{Bramel}{D.}
\AuthorIndex{Chae}{E.}
\AuthorIndex{Covault}{C.\ E.}
\AuthorIndex{Fortin}{P.}
\AuthorIndex{Gingrich}{D.\ M.}
\AuthorIndex{Hanna}{D.\ S.}
\AuthorIndex{Hinton}{J.\ A.}
\AuthorIndex{Mukherjee}{R.}
\AuthorIndex{Mueller}{C.}
\AuthorIndex{Ong}{R.\ A.}
\AuthorIndex{Ragan}{K.}
\AuthorIndex{Scalzo}{R.\ A.}
\AuthorIndex{Schuette}{D.\ R.}
\AuthorIndex{Th\'eoret}{C.\ G.}
\AuthorIndex{Williams}{D.\ A.}


\section*{Abstract}

STACEE is a ground based $\gamma$-ray observatory that uses a heliostat
array, built for solar energy research, to detect atmospheric Cherenkov
radiation from $\gamma$-ray initiated extensive air showers.  During the
first half of 2001, a prototype detector, STACEE-48, was used to detect
the blazar Markarian 421, which was in an extremely active state.
Observations from March to May of 2001 yielded an integral flux of
$(8.0\pm0.7_{stat}\pm1.5_{sys})\times 10^{-10} {\rm cm}^{-2}{\rm s}^{-1}$
at energies above $140\pm20$ GeV, and provide some evidence of
correlated trends on time scales of a week or more in the GeV and X-ray
bands.

\section{Introduction}


The STACEE--48 detector (Covault et al.\ 2001, Boone 2002) operated from
2000 to 2001 at the National Solar Thermal Test Facility in New Mexico,
USA.  This prototype of the full STACEE experiment employed 48 steerable
mirrors (heliostats), distributed over approximately $2\times 10^4$
square meters.  This large collection area allows STACEE to operate into the
lower energy $\gamma$-ray regime left unexplored by previous ground-based
techniques.

Science observations of Markarian 421 were conducted from March through
May of 2001, constituting 26 nights and roughly 36 hours of exposure time.
Of this, 22 hours of source observations
were deemed good-quality data and used for this analysis.  Simulations of
the detector performance indicate that the peak of the STACEE-48 energy
response to a source with the approximate spectral index of Markarian 421
occurs at $140\pm 20$ GeV, which has been adopted as our energy threshold
for these data (see Boone et al.\ 2002 for details).  Thus, these
observations represent the lowest energy $\gamma$-ray detection of the
2001 flare by a ground-based instrument.


\section{Spectral Results}

The STACEE observations of Markarian 421 yielded an average
$\gamma$-ray rate of $7.7\pm0.7\pm1.2$ min$^{-1}$.  Included in this rate
is a correction for false triggers due to the presence of a bright star
(HD 95934, magnitude 6.16 in the B band) within a few arcminutes of the
source.  The correction was determined by observing a star of comparable
brightness (HIP 80460) with no known associated $\gamma$-ray source in
the field of view.  The excess rate from the HIP 80460 observations was
attributed to a constant star effect, and thus subtracted from the
Markarian 421 observations.  The above rate, assuming a differential
spectral index of $\alpha = 2.1$, corresponds to an average integral
$\gamma$-ray flux above 140 GeV of
$$
\Phi(E>140\ {\rm GeV}) = (8.0\pm0.7_{stat}\pm1.5_{sys})\times 10^{-10}
{\rm cm}^{-2}{\rm s}^{-1}.
$$

\begin{figure}[t]
   \begin{center}
   \includegraphics[width=17pc]{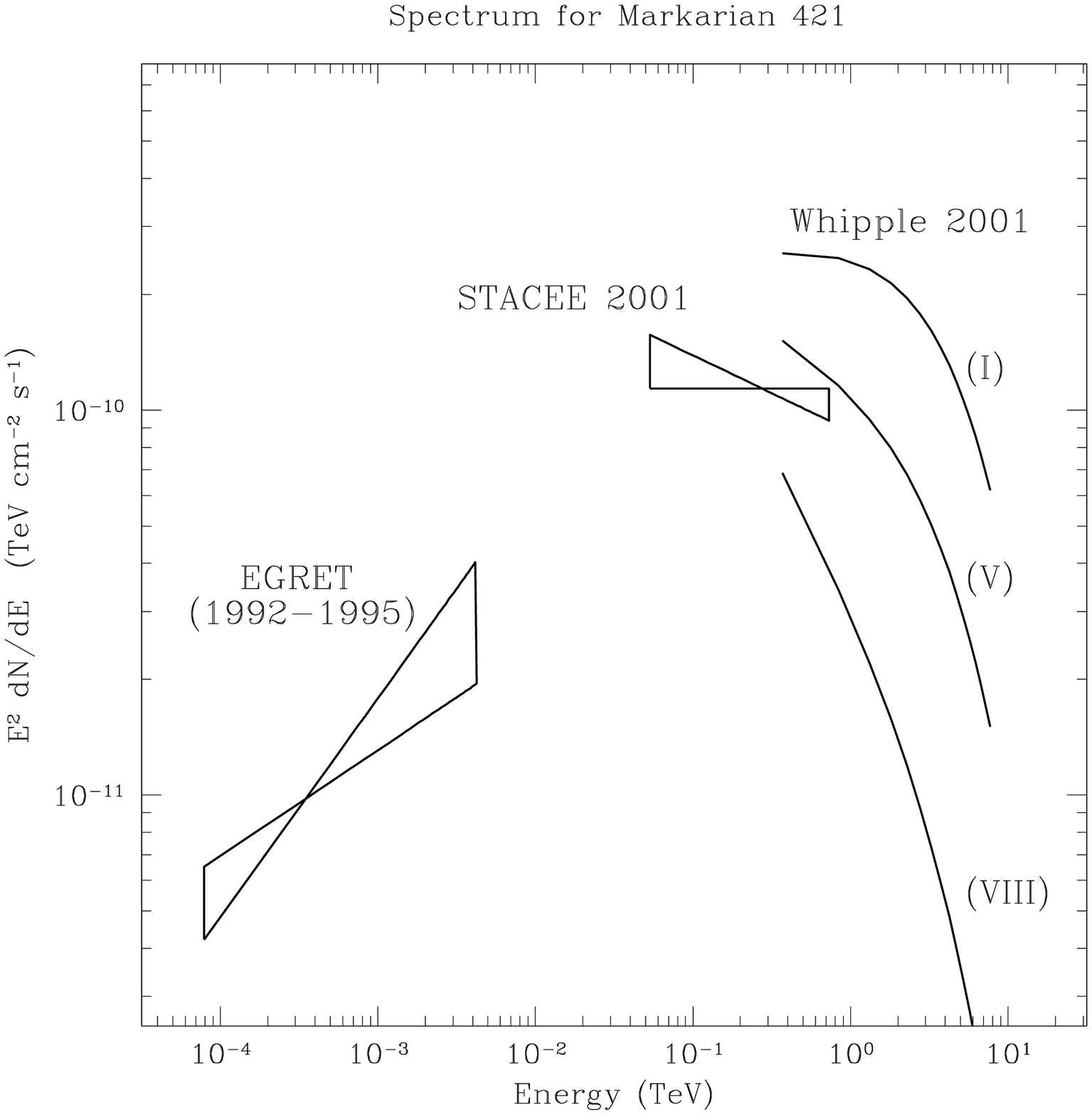}
   \includegraphics[width=17pc]{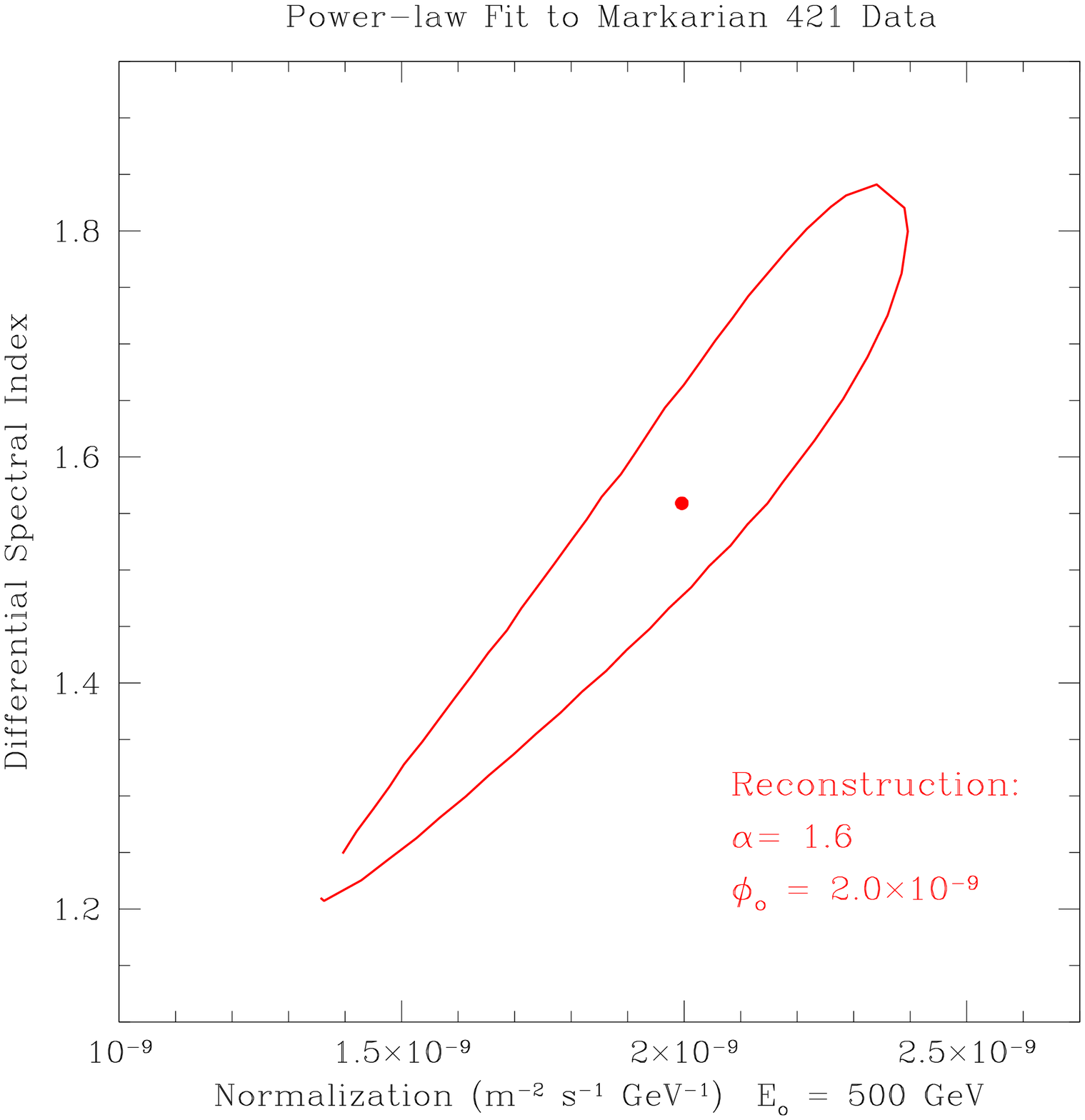}
   \end{center}
   \caption{{\bf a)} Spectral energy distribution for Markarian 421.
      The STACEE butterfly corresponds to the average flux for the 2001
      flare, while the Whipple curves represent different flaring states
      from the same time frame.  The EGRET butterfly is included for
      reference.
      {\bf b)} Forward-folding fit of a simple power-law spectrum to
      the  STACEE observations of Markarian 421 in Spring of 2001.
      The Y axis is the differential spectral index, while the X axis
      is the normalization. The contour is the $1\sigma$ error contour.
   }
\end{figure}

If the spectrum of Markarian 421 is assumed to follow a power-law form in
the range of energies to which STACEE is sensitive, the normalization of
the differential spectrum can be inferred from the measured rate,
provided the spectral index is assumed.  In Fig.~1a (Boone et al.\ 2002),
the range of possible power-law solutions to the STACEE rate for
differential indices between 2.00 and 2.20 is given by the
butterfly-shaped region.
For comparison, Whipple observations around the same time are plotted
for high (I), medium (V), and low (VIII) flux states
(Krennrich et al.\ 2002).  EGRET data from cycles 1 through 5 are also
plotted for reference (Hartman et al.\ 1999).

Since STACEE--48 did not record pulse amplitude information for each
event, an
unbiased determination of the differential flux is difficult.  However,
because the energy response of STACEE changes significantly as a function
of the source elevation and the trigger configuration, it is possible to
glean some information about the differential spectrum by using the
forward-folding technique.
Under this prescription, the Markarian 421 data set was first divided into
elevation bands.  The simulated detector response to an assumed power-law
spectrum at a given elevation was then compared to the appropriate subset
of the Markarian 421 observations for several different reimposed trigger
criteria.  This process was repeated for a number of different elevation
bands.  The differential index ($\alpha$) and the normalization ($\phi_0$)
of the assumed spectrum were then adjusted to minimize the difference
between the expected and observed responses over all considered
combinations.

\begin{figure}[t]
   \begin{center}
   \includegraphics[width=17pc]{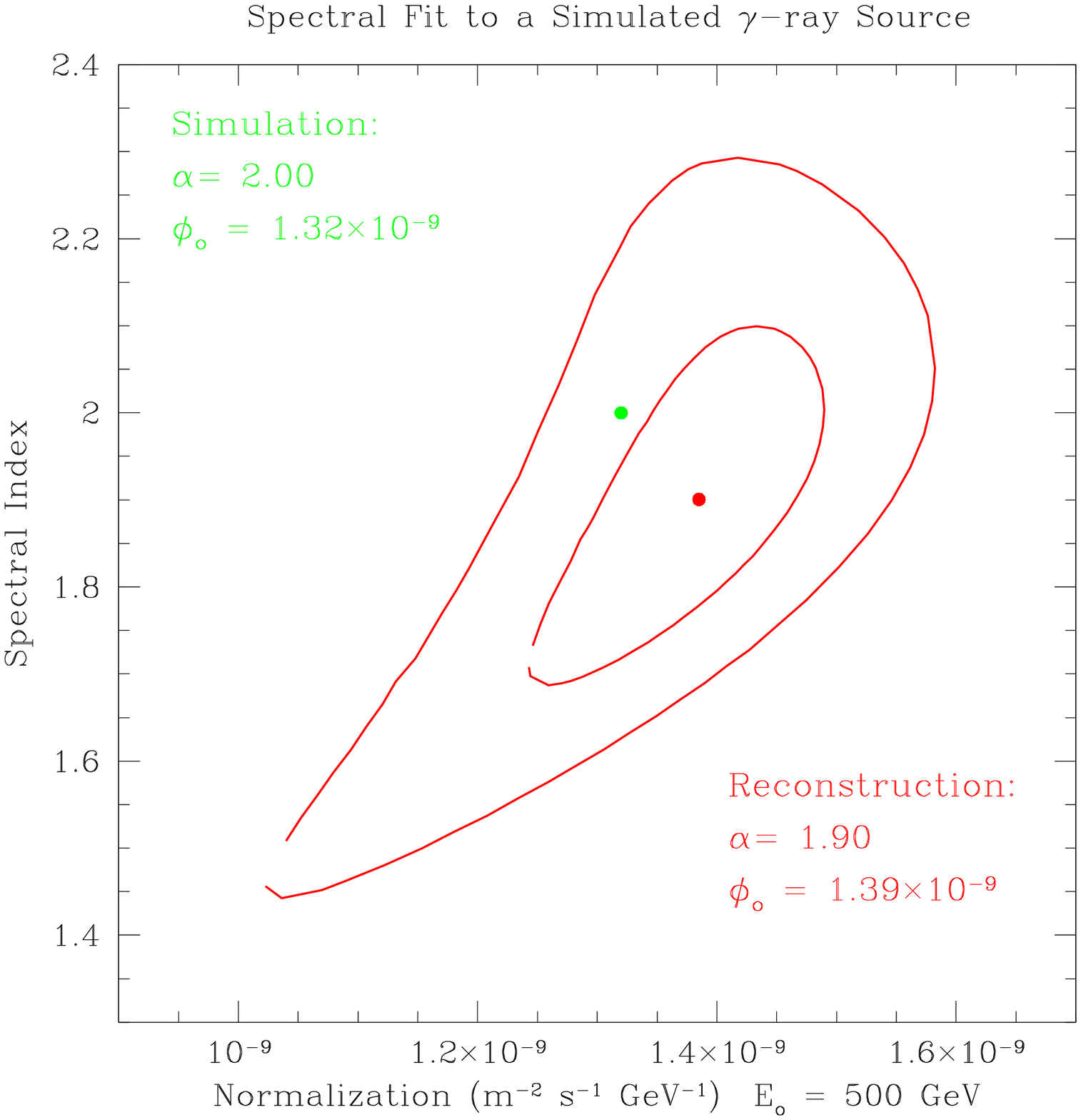}
   \includegraphics[width=17pc]{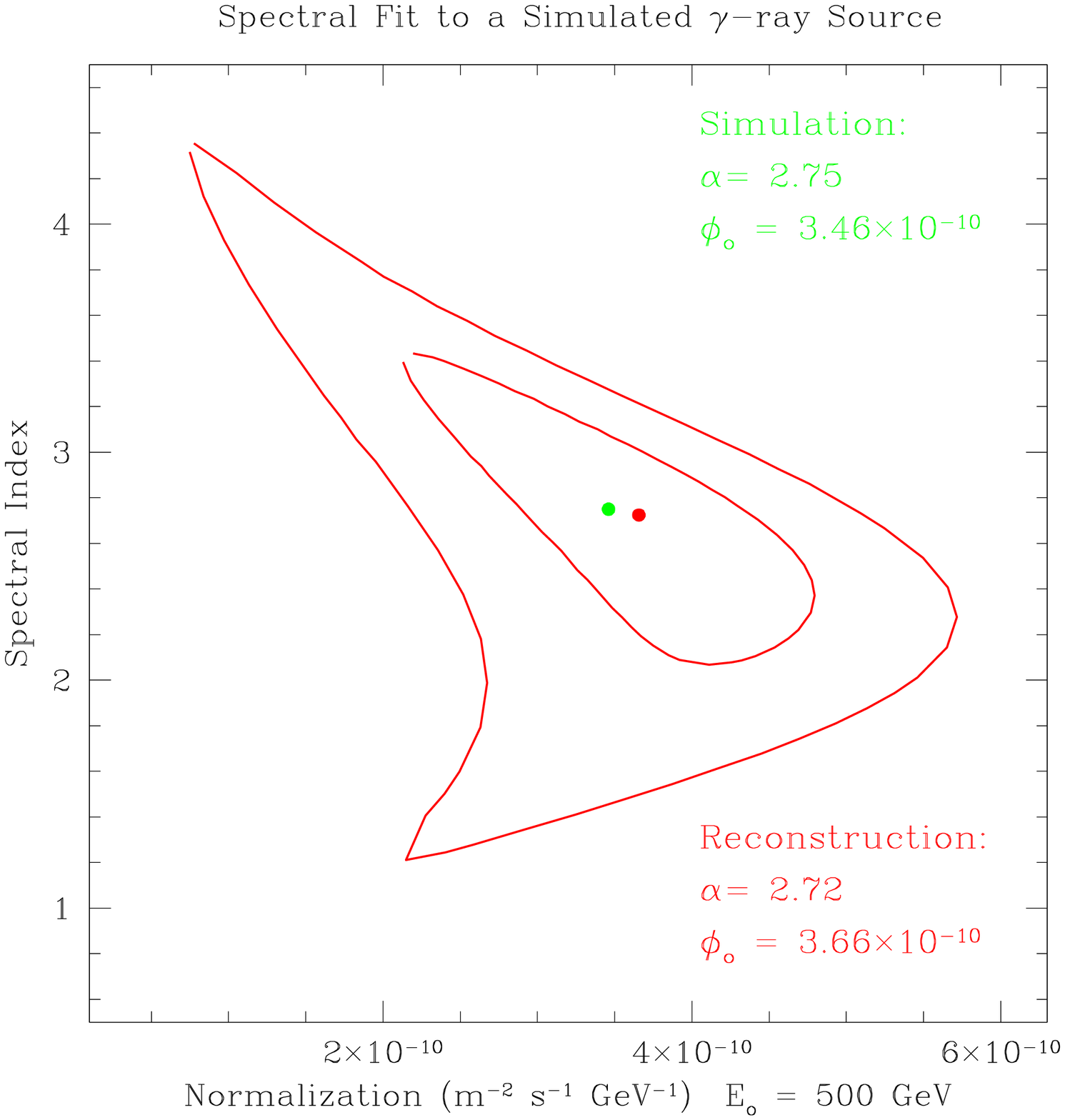}
   \end{center}
   \caption{Forward folding fit results for simulated source spectra.
      {\bf a)} Results for a simulated source of differential spectral
      index $\alpha=2.00$.  {\bf b)} Results for a simulated source of
      differential spectral index $\alpha=2.75$.}
\end{figure}

Figure 2 includes two tests of this method on simulated data sets with
integral fluxes approximately ten times that expected from Markarian 421
(contours represent
one sigma errors).  Note that the normalization energy ($E_0$) has been
adjusted to the non-standard value of 500 GeV.  This transformation was
necessary to decrease the strong correlation in the two parameters.
%
%
%
The success of the simulated tests indicates that, in principle, the
technique is capable of extracting the spectral parameters.
%
%
%
%
However, due to systematic effects in the simulations themselves, these
tests do not vouch for the validity of the technique when used on actual
data.

This discrepancy is reflected in the application of the technique to the
Markarian 421 data in Fig.~1b (contour is the one sigma error).  In
units of TeV, the reconstructed normalization translates to
$(6.8\pm1.7)\times10^{-11}$ cm$^{-2}$ s$^{-1}$ TeV$^{-1}$.  This value, in
conjunction with the extremely hard spectral index of 1.6 does not agree
with the average STACEE flux in Fig.~1a or the STACEE/Whipple
flux ratios described in Boone et al.\ (2002).  Whether these
inconsistencies arise from inaccuracies in our detector simulations, the
inappropriateness of a power-law postulate, insufficient statistics, or
the high variability of the source, remains to be seen.


\section{Temporal Results}

Figure 3a (Boone et al.\ 2002) contains three light curves for Markarian
421 from March through May of 2001.  The upper panel, from the RXTE All Sky
Monitor is from the 2 to 20 keV energy band (results provided by the RXTE
ASM team).  The lower panel, from the Whipple detector
(Holder et al.\ 2001), spans energies from 250 GeV to 8 TeV.  The STACEE
observations, depicted in the central panel, represent the photon flux
between 50 and 300 GeV.  Points are photon count rates, averaged on a day
by day basis.

\begin{figure}[t]
   \begin{center}
   \includegraphics[width=17pc]{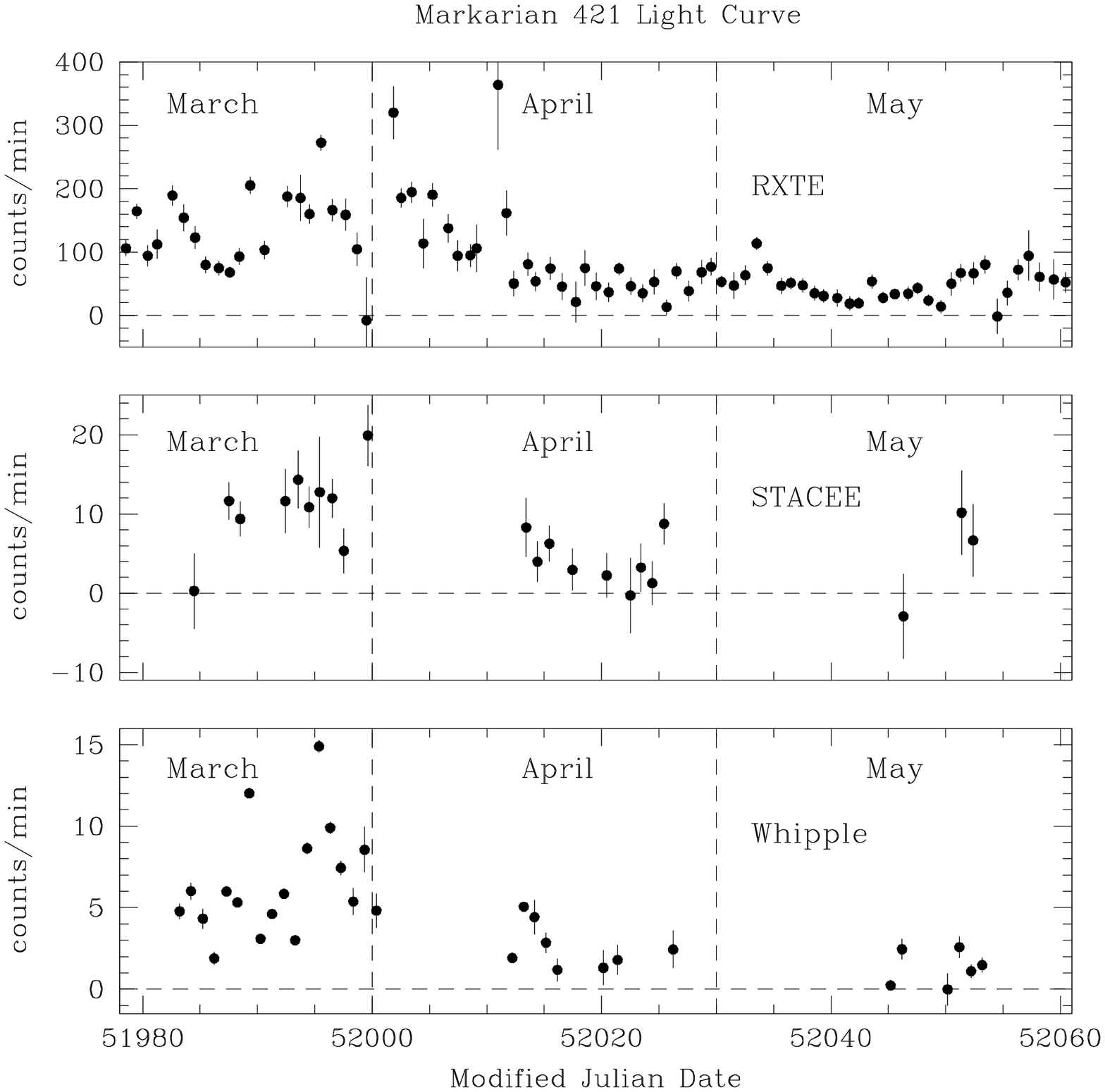}
   \includegraphics[width=17pc]{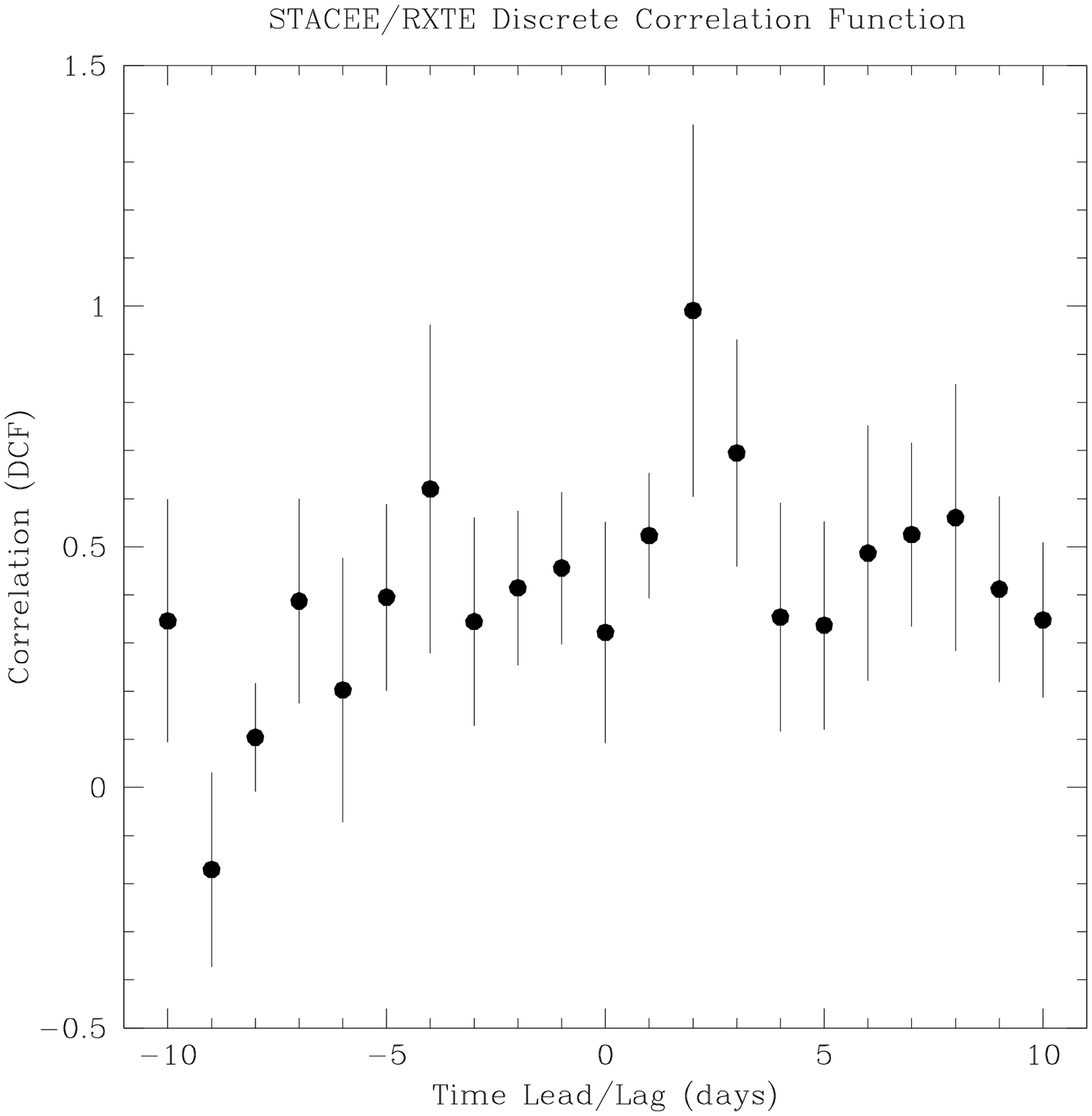}
   \end{center}
   \caption{{\bf a)} Multiwavelength lightcurve for Markarian 421 during
      the 2001 flare.  Panels are, from top to bottom, RXTE data, STACEE
      data, and Whipple data.  References cited in text.
      {\bf b)} Correlation analysis of STACEE and RXTE data using the
      discrete correlation function.
   }
\end{figure}

A correlation analysis was performed on the STACEE and RXTE data sets
using the discrete correlation function (DCF) described in
Edelson \& Krolik (1988).  Unbinned discrete correlation coefficients
(UDCF) were formed for each pair of points in the respective data sets as
follows:
$$
{\rm UDCF}_{ij}={({\rm STACEE}_i-\overline{\rm STACEE})({\rm RXTE}_j -
\overline{\rm RXTE}) \over\sigma_{\rm STACEE}\sigma_{\rm RXTE}}.
$$
These coefficients were then grouped by their associated time differences
($\Delta t=t_{\rm RXTE} - t_{\rm STACEE}$), and averaged to form the DCF.

Figure 3b is a plot of the DCF as a function of time lag for the STACEE
and RXTE data sets between MJD 51984 and MJD 52052.  Error bars are
constructed according to the DCF prescription, and can be understood as
the error on the mean for each bin.  Although the peak value of the DCF
occurs at two days (corresponding to the STACEE data leading the RXTE
data by two days), the large error bars on this point indicate that it is
not statistically significant with respect to a value of zero (only
2.6$\sigma$).  The most significant point in the DCF is $\Delta t=1$
(4.0$\sigma$). However, as most bins have approximately 20 entries, it is
expected that even completely uncorrelated data sets could exhibit DCF
values of $\sim0.2$ (see Edelson \& Krolik 1988 for details). Thus, it is
difficult to draw any meaningful conclusions from the DCF alone.


\section{Conclusions}

The STACEE--48 instrument was used to successfully detect Markarian 421
during its flare in the early part of 2001.  The average flux observed
appears consistent with the variable spectral energy distribution (SED)
reported by the Whipple experiment.  The data are not suitable for an
event-by-event energy analysis, and the forward folding technique yields
results which are not consistent with the STACEE average flux. Therefore,
it appears that the spectral analysis of these data may be limited to the
time-averaged flux reported here and elsewhere. However, the SED shown in
Fig.~1a suggests that the STACEE sensitivity may include the peak of the
TeV emission for Markarian 421.  Thus, more detailed spectral information
from STACEE is highly desirable.  The current incarnation of the
experiment, STACEE--64, is instrumented with waveform digitizers on all
channels, which should greatly improve the energy resolution.

A visual inspection of the contemporaneous light curves from RXTE, STACEE,
and Whipple suggests correlated trends across all three energy bands. And
though a more formal correlation analysis with the DCF does not yet
provide compelling statistical evidence for a correlation on day time
scales, it does not preclude it either.  We suspect that this ambiguity
may be due to the rather sparse sampling in the STACEE data set.  In
fact, it may be that the traditional DCF analysis is generally less
suited to GeV-TeV observations, which tend to redefine the notion of
``sparsely sampled'' data.  Alternative timing analysis methods are
currently under investigation.


\section{References}

\re 1.\ Boone, L.~M., et al.\ 2002, ApJ, 579, L5
\re 2.\ Boone, L.~M. 2002, PhD. thesis, Univ. California, Santa Cruz
\re 3.\ Covault, C.~E., et al.\ 2001, in Proc 27th ICRC (Hamburg), 1, 2810
\re 4.\ Edelson, R.~A.~\& Krolik, J.~H., 1988, ApJ, 333, 646
\re 5.\ Hartman, R.~C., et al.\ 1999, ApJS, 123, 79
\re 6.\ Holder, J., et al.\ 2001, in Proc 27th ICRC (Hamburg), 1, 2613
\re 7.\ Krennrich, F., et al.\ 2002, ApJ, 575, L9

\endofpaper
\end{document}